\documentclass[12pt,preprint]{aastex}

\newcommand{\eqw}{\ensuremath{W_{\lambda}}}
\newcommand{\h}{\ensuremath{\,h_{70}^{-1}}}
\newcommand{\hsq}{\ensuremath{\,h_{70}^{-2}}}
\newcommand{\Inu}{\ensuremath{I_{\nu}}}
\newcommand{\kms}{{\rm km~s}\ensuremath{^{-1}}}
\newcommand{\LB}{\ensuremath{L_B}}
\newcommand{\Lstar}{\ensuremath{\,L^*}}
\newcommand{\MB}{\ensuremath{M_B}}
\newcommand{\persqarcsec}{\ensuremath{{\rm arcsec^{-2}}}}
\newcommand{\persqcm}{\ensuremath{{\rm cm^{-2}}}}
\newcommand{\percubcm}{\ensuremath{{\rm cm^{-3}}}}
\newcommand{\perHz}{\ensuremath{{\rm Hz^{-1}}}}
\newcommand{\percubpc}{\ensuremath{{\rm pc^{-3}}}}
\newcommand{\persqkpc}{\ensuremath{{\rm kpc^{-2}}}}
\newcommand{\persec}{\ensuremath{{\rm s^{-1}}}}
\newcommand{\peryr}{\ensuremath{{\rm yr^{-1}}}}
\newcommand{\Zsun}{\ensuremath{Z_{\Sun}}}

\newcommand{\czabs}{\ensuremath{cz_{\rm abs}}}
\newcommand{\czgal}{\ensuremath{cz_{\rm gal}}}

\newcommand{\vesc}{\ensuremath{v_{\rm esc}}}
\newcommand{\vlos}{\ensuremath{v_{\rm los}}}
\newcommand{\vw}{\ensuremath{v_{\rm w}}}

\newcommand{\MHI}{\ensuremath{{M_{\rm H\,I}}}}
\newcommand{\Mdyn}{\ensuremath{{M_{\rm dyn}}}}
\newcommand{\Msun}{\ensuremath{{M_{\Sun}}}}

\newcommand{\sbs}{SBS~1122+594}
\newcommand{\ic}{IC~691}
\newcommand{\sdsslsb}{SDSS~J1126+593}

\newcommand{\Lya}{{\rm Ly}\ensuremath{\alpha}}
\newcommand{\Ha}{{\rm H}\ensuremath{\alpha}}
\newcommand{\Hb}{{\rm H}\ensuremath{\beta}}

\newcommand{\AlII}{\ion{Al}{2}}
\newcommand{\CII}{\ion{C}{2}}
\newcommand{\CIII}{\ion{C}{3}}
\newcommand{\CIV}{\ion{C}{4}}
\newcommand{\HI}{\ion{H}{1}}
\newcommand{\HII}{\ion{H}{2}}
\newcommand{\MgII}{\ion{Mg}{2}}
\newcommand{\NII}{\ion{N}{2}}
\newcommand{\OI}{\ion{O}{1}}
\newcommand{\OII}{\ion{O}{2}}
\newcommand{\OIII}{\ion{O}{3}}
\newcommand{\OVI}{\ion{O}{6}}
\newcommand{\SII}{\ion{S}{2}}
\newcommand{\SiII}{\ion{Si}{2}}
\newcommand{\SiIII}{\ion{Si}{3}}
\newcommand{\SiIV}{\ion{Si}{4}}

\newcommand{\NHI}{\ensuremath{N_{\rm H\,I}}}
\newcommand{\NCIV}{\ensuremath{N_{\rm C\,IV}}}

\begin{document}

\title{Discovery of a Metal-Line Absorber Associated with a Local Dwarf Starburst Galaxy}
\author{Brian A. Keeney\altaffilmark{1}, John T. Stocke\altaffilmark{1}, Jessica L. Rosenberg\altaffilmark{2}, Jason Tumlinson\altaffilmark{3}, \\and Donald G. York\altaffilmark{4,5}}
\altaffiltext{1}{Center for Astrophysics and Space Astronomy, Department of Astrophysical and Planetary Sciences, University of Colorado, 389 UCB, Boulder, CO 80309; keeney@casa.colorado.edu, stocke@casa.colorado.edu}
\altaffiltext{2}{NSF Astronomy and Astrophysics Postdoctoral Fellow, Harvard-Smithsonian Center for Astrophysics, 60 Garden Street, Cambridge, MA 02138; jlrosenberg@cfa.harvard.edu}
\altaffiltext{3}{Yale Center for Astronomy and Astrophysics, Departments of Physics and Astronomy, Yale University, New Haven, CT 06520; jason.tumlinson@yale.edu}
\altaffiltext{4}{Department of Astronomy and Astrophysics, University of Chicago, 5640 South Ellis Avenue, Chicago, IL 60637}
\altaffiltext{5}{Enrico Fermi Institute, University of Chicago, 5640 South Ellis Avenue, Chicago, IL 60637}

\shorttitle{Absorber Associated with Dwarf}
\shortauthors{Keeney et al.}

\begin{abstract}

We present optical and near-infrared images, \HI\ 21~cm emission maps, optical spectroscopy, and {\em Hubble Space Telescope}/Space Telescope Imaging Spectrograph ultraviolet spectroscopy of the QSO/galaxy pair \sbs/\ic. The QSO sight line lies at a position angle of $27\degr$ from the minor axis of the nearby dwarf starburst galaxy \ic\ ($\czgal = 1204\pm3$~\kms, $\LB \sim 0.09$~\Lstar, current star formation ${\rm rate} = 0.08$--0.24\hsq~\Msun~\peryr) and 33\h~kpc ($6\farcm6$) from its nucleus. We find that \ic\ has an \HI\ mass of $\MHI = (3.6\pm0.1)\times10^8$~\Msun\ and a dynamical mass of $\Mdyn = (3.1\pm0.5)\times10^{10}\h$~\Msun. The UV spectrum of \sbs\ shows a metal-line (\Lya+\CIV) absorber near the redshift of \ic\ at $\czabs = 1110\pm30$~\kms. Since \ic\ is a dwarf starburst and the \sbs\ sight line lies in the expected location for an outflowing wind, we propose that the best model for producing this metal-line absorber is a starburst wind from \ic. We place consistent metallicity limits on \ic\ ($[Z/\Zsun] \sim -0.7$) and the metal-line absorber ($[Z/\Zsun] < -0.3$). We also find that the galaxy's escape velocity at the absorber location is $\vesc = 80\pm10$~\kms\ and derive a wind velocity of $\vw = 160\pm50$~\kms. Thus, the evidence suggests that \ic\ produces an unbound starburst wind that escapes from its gravitational potential to transport metals and energy to the surrounding intergalactic medium.

\end{abstract}


\section{Introduction}
\label{intro}

Metals have been distributed throughout much of the intergalactic medium (IGM) at all observed redshifts. Transport by galactic starburst winds is a leading theory for explaining the existence of these IGM metals, which are often found far \citep[several hundred kiloparsecs;][]{stocke06b} from any sites of ongoing star formation. However, it is not clear whether starbursts associated with luminous massive galaxies or smaller but more numerous dwarf galaxies produce the winds that are primarily reponsible for enriching the IGM with metals and energy. Statistically, it appears that 0.1~\Lstar\ galaxies must contribute to IGM enrichment unless \Lstar\ galaxies can enrich regions up to $\sim 1$~Mpc in radius \citep{tumlinson05a}.

Starburst winds have been studied around nearby galaxies in emission using narrowband nebular emission line (e.g., \Ha) and thermal X-ray continuum images \citep*{watson84,martin99,martin02} as well as ultraviolet and optical absorption-line spectroscopy \citep*{heckman00,heckman01,rupke05a,rupke05b,martin05,martin06,keeney05,keeney06}. Both imaging and absorption-line spectroscopy have limitations, but the two techniques are complementary in many ways. Emission-line studies can address the extent and morphology of the outflow, but only in its densest regions. These studies cannot measure the velocity or temperature of the outflow in the diffuse halo, so they are incapable of determining whether ejecta in the halo are bound to the galaxy. Thus, emission-line studies alone cannot determine which galaxies produce outflows that enrich the surrounding IGM. Absorption-line studies, on the other hand, are much more sensitive to diffuse gas and therefore can determine whether ejecta in the halo are bound to the galaxy, but only along one line of sight so they are unable to study the outflow morphology.

Most absorption-line studies also introduce an ambiguity in the distance between the background source and the absorbing gas since they typically use the stellar continuum of the starburst itself as their source \citep[e.g.,][]{heckman00,heckman01,rupke05a,rupke05b,martin05,martin06}. Consequently, they cannot distinguish a high-velocity outflow in the galactic halo from one in the starburst region itself. Using outflow velocities measured in the galactic disk to predict whether the absorbing gas is gravitationally bound to the galaxy is problematic because ejecta within several kiloparsecs of the starburst region will decelerate under the combined effects of gravity and mass loading \citep[e.g.,][]{suchkov96,martin02,strickland04}. This problem can be circumvented by probing the outflow well away from the starburst region, which is only feasible if there is a bright background QSO projected near the galaxy of interest \citep[e.g.,][]{stocke04,keeney05,keeney06}.

Massive galaxies have high rates of metal production and thus large amounts of potential fuel for IGM enrichment. Luminous low-redshift starbursts have star formation rates (SFRs) in the range 1--10~\Msun~\peryr, although rates of $\sim 100$--1000~\Msun~\peryr\ can be triggered by mergers \citep[\citealp*{heckman90};][]{martin03,rupke05a,rupke05b}. These galaxies generate winds with average outflow velocities of 300--400~\kms, but wind speeds of $\gtrsim 1000$~\kms\ are not unheard of \citep*{heckman00,filippenko92,veilleux94,veilleux05}. Only the most luminous galaxies can be detected at high redshift, so studies of the earliest epochs of star formation and metal enrichment are inherently biased toward the most massive starbursts \citep*{steidel96a,steidel96b,steidel99,steidel01,pettini01,pettini02,adelberger03}.

Dwarf galaxies produce fewer metals per galaxy than their massive counterparts but they are much more numerous and their cumulative effects could dominate IGM enrichment \citep{stocke04,tumlinson05a}. Dwarf starburst galaxies have typical SFRs of 0.1--1~\Msun~\peryr\ \citep{martin03} and produce winds with outflow velocities of $\sim 50$--200~\kms\ \citep{marlowe95,martin98,schwartz04}. However, starbursts of all luminosities have the same maximum areal SFR \citep[$\sim 45$~\Msun~\persqkpc~\peryr;][]{meurer97} and produce winds with the same X-ray temperature \citep{martin99}. These results indicate that dwarf starbursts may be more efficient than their massive counterparts at transporting the metals entrained in their winds to the IGM due to their shallower gravitational potentials with lower escape velocities.

Recent wind studies that utilize nearby QSOs to probe starburst winds several kiloparsecs from the host galaxy find that luminous galaxies produce bound winds while dwarf galaxies produce unbound winds. \citet{keeney05} found that the nearby luminous starburst galaxy NGC~3067 ($L \approx 0.5$~\Lstar, ${\rm SFR} \approx 1.4$~\Msun~\peryr) produces a bound wind along the line of sight to 3C~232, which is located near the minor axis of NGC~3067 and 11\h~kpc from the galactic plane. The Milky Way also produces a bound starburst wind along the lines of sight to two high-latitude AGN (Mrk~1383 and PKS~2005--489) at $l=350\degr$ that probe the regions on either side of the Galactic Center at heights up to 12.5~kpc \citep{keeney06}. On the other hand, \citet{stocke04} found that a dwarf poststarburst galaxy produced an unbound wind $\sim 3.5$~Gyr ago that is now observed as the 1586~\kms\ metal-line absorber in the spectrum of 3C~273, which is 71\h~kpc away in projection on the sky.

In our on-going study to determine what type of galaxies enrich the IGM, this paper presents an ultraviolet spectrum of the QSO \sbs, as well as optical and near-infrared images, an \HI\ 21~cm emission map, and an optical spectrum of the blue compact galaxy \ic, which is $6\farcm6$ away on the sky. These observations will be used to study whether the starburst wind produced by \ic\ can escape its gravitational potential to enrich the surrounding IGM. In \S\,\ref{obs} we describe the acquisition and reduction of our multi-wavelength observations. The connection between the metal-line absorber in the spectrum of \sbs\ and \ic\ is discussed in \S\,\ref{galabs}.  In \S\,\ref{escape} we examine whether the starburst wind produced by \ic\ can escape from its gravitational potential. We summarize our results and discuss their implications for IGM enrichment in \S\,\ref{conclusion}.

\section{Observations and Data Reduction}
\label{obs}

We have obtained images and spectra of the QSO/galaxy pair \sbs/\ic\ at several wavelengths. Our dataset includes ultraviolet and optical spectra of \sbs, as well as optical, near-infrared, and \HI\ 21~cm images and a long-slit optical spectrum of \ic. The acquisition and reduction of these data are described below in \S\S\,\ref{obs:qsospec}--\ref{obs:galemmap}. Broadband optical images and fiber spectra of \sbs\ and \ic\ are also available from the Sloan Digital Sky Survey \citep[SDSS;][]{york00,abazajian05}. \sbs\ is located $6\farcm6$ from \ic, which corresponds to an impact parameter of 33\h~kpc at the redshift of \ic\ assuming a distance of $\czgal/H_0 = 17.2$\h~Mpc ($\czgal = 1204\pm3$~\kms; see \S\,\ref{obs:galemmap}).

\subsection{Ultraviolet and Spectrum of \sbs}
\label{obs:qsospec}

\sbs\ was observed by the {\em Hubble Space Telescope} (HST) with the G140L grating of the Space Telescope Imaging Spectrograph (STIS) for 1320~seconds on 2004 Apr 6 as part of GO Program 9874 (PI: J. Tumlinson). Despite the short exposure time of this snapshot spectrum, an intergalactic \Lya+\CIV\ absorber is evident near the redshift of \ic. Figure~\ref{fig:qsospec} shows the associated \Lya\ and \CIV\ absorption features, as well as an intergalactic \Lya\ line at $cz \approx 2350$~\kms\ that can be seen in the upper panel. Information about other features in this spectrum, which covers wavelengths of $\sim 1120$--1720~\AA, can be found in the Appendix.

The continuum near the associated \Lya\ and \CIV\ absorption lines was normalized with Legendre polynomials and the \Lya\ and \CIV\ features in the normalized spectra were fitted with Voigt profiles using a $\chi^2$ minimization routine developed by B.A.K. Rest wavelengths, oscillator strengths, and transition rates for the Voigt profile fits were taken from \citet{morton03}. The fit to the \CIV\ data was constrained such that both lines in the doublet \citep[rest wavelengths of 1548.2 and 1550.8~\AA;][]{morton03} were fitted with the same velocity, Doppler $b$ value, and ionic column density, resulting in a best fit with  a velocity of $\czabs = 1110\pm30$~\kms\ and a rest-frame equivalent width in the stronger, bluer line of $800\pm200$~m\AA. The best-fit to the intergalactic \Lya\ absorber found a velocity of $1250\pm70$~\kms\ and a rest-frame equivalent width of $1700\pm200$~m\AA. However, this absorber is located on the wing of the Galactic \Lya\ line and is further blended with another intergalactic \Lya\ absorber at $cz \approx 2350$~\kms\ at the low resolution of STIS with the G140L grating (see Fig.~\ref{fig:qsospec}). We believe that this blending and the poorer resolution in the \Lya\ region ($\approx 430$~\kms) as compared to the \CIV\ region ($\approx 340$~\kms) make the best-fit \Lya\ velocity untrustworthy. Therefore, we adopt the best-fit \CIV\ velocity of $\czabs = 1110\pm30$~\kms\ as the velocity of the \Lya+\CIV\ absorber. This velocity is indicated by a vertical gray bar in Figure~\ref{fig:qsospec}.

We have estimated the \Lya\ and \CIV\ column densities for this absorber using the apparent optical depth (AOD) method \citep{savage91}. This method predicts a \HI\ column density of $\NHI = 10^{14.6\pm0.1}$~\persqcm\ for this absorber, which we treat as a lower limit to the true column density since the AOD method underpredicts the column density of saturated lines. Unresolved saturation is likely in this \Lya\ absorber due to the low velocity resolution of our G140L spectrum (e.g., the trough of the Galactic \Lya\ absorption feature in Fig.~\ref{fig:qsospec} is well above zero). Both lines of the \CIV\ doublet yield AOD column densities that agree to within the combined errors. We adopt the column density predicted by the weaker line of the doublet, $\NCIV = 10^{14.7\pm0.2}$~\persqcm, since it is less susceptible to unresolved saturation. These column densities are used to estimate the absorber metallicity in \S\,\ref{galabs}.

\subsection{Optical and Near-Infrared Images of \ic}
\label{obs:galimg}

\begin{figure}[!ht]
\begin{center}
\epsscale{1.00}
\plotone{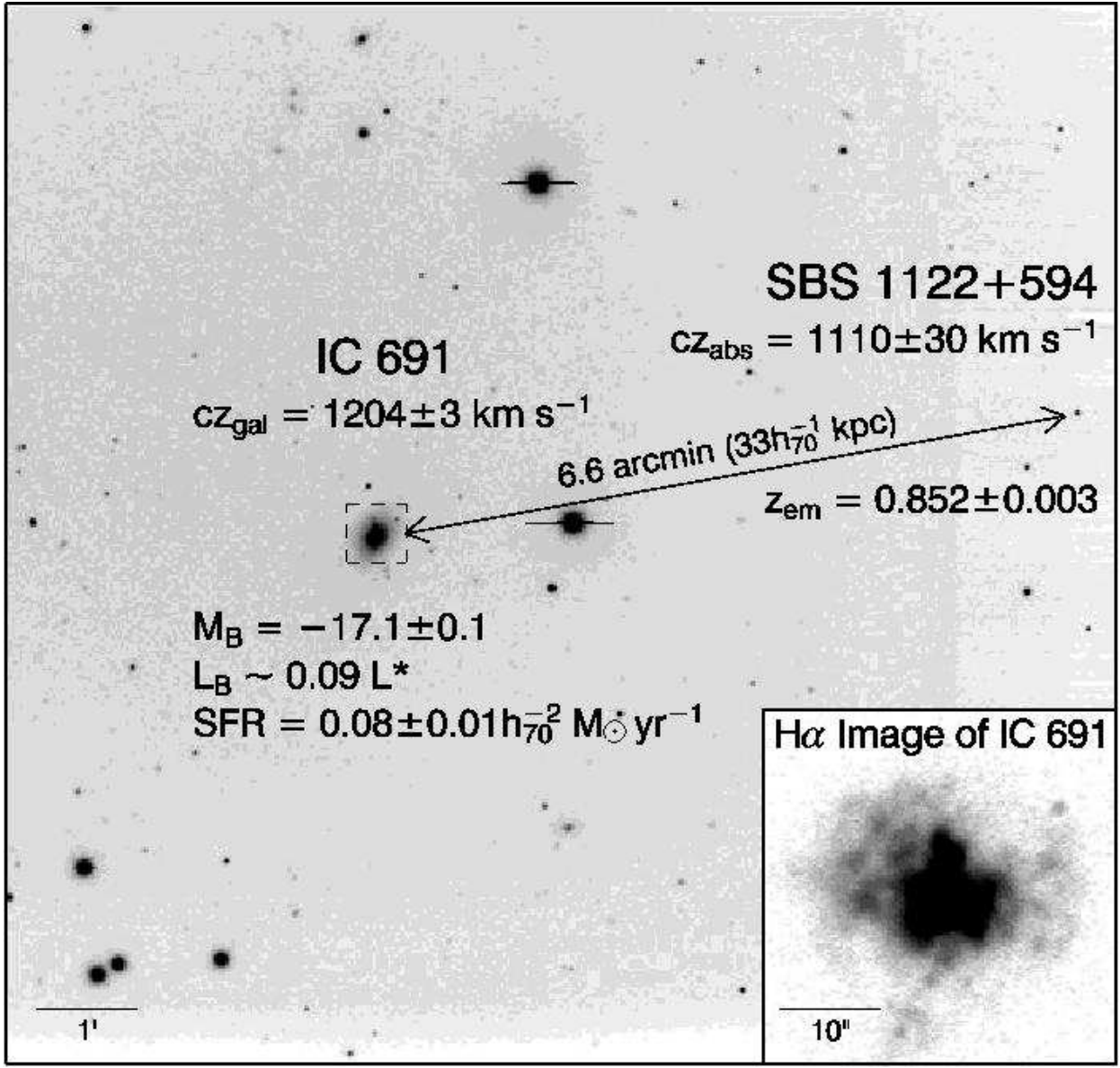}
\end{center}
\caption{R-band mosaic of the QSO/galaxy pair \sbs/\ic\ taken with the KPNO 2.1-m telescope. The inset, whose approximate boundary is marked by the dashed box in the mosaic, shows a pure \Ha\ image of \ic. The seeing in the mosaic is $1\farcs8$ and the seeing in the inset is $1\farcs5$.
\label{fig:Halpha}}
\end{figure}

Broadband optical and \Ha\ images of \ic\ were obtained with the T2KA CCD at the Kitt Peak National Observatory (KPNO) 2.1-m telescope on 2003 Mar 4 and 2003 Mar 6. \ic\ was observed for a total exposure time of 480~s in B-band, 1680~s in R-band, and 2400~s each in the \Ha\ on- and off-band filters. A mosaic of R-band images showing both \ic\ and \sbs\ is shown in Figure~\ref{fig:Halpha}. Relative photometry with in-field standard stars yields an apparent B-band magnitude for \ic, integrated to the 25th~mag~\persqarcsec\ isophote ($r_{25} = 39\arcsec = 3.3\h$~kpc), of $m_{B,25} = 14.1\pm0.1$, which corresponds to an absolute magnitude of $\MB = -17.1\pm0.1$ and a luminosity of $\LB \approx 0.09$~\Lstar\ \citep*[using $\MB^*$ from][]{marzke94}. 

\begin{figure}[!ht]
\begin{center}
\epsscale{1.00}
\plotone{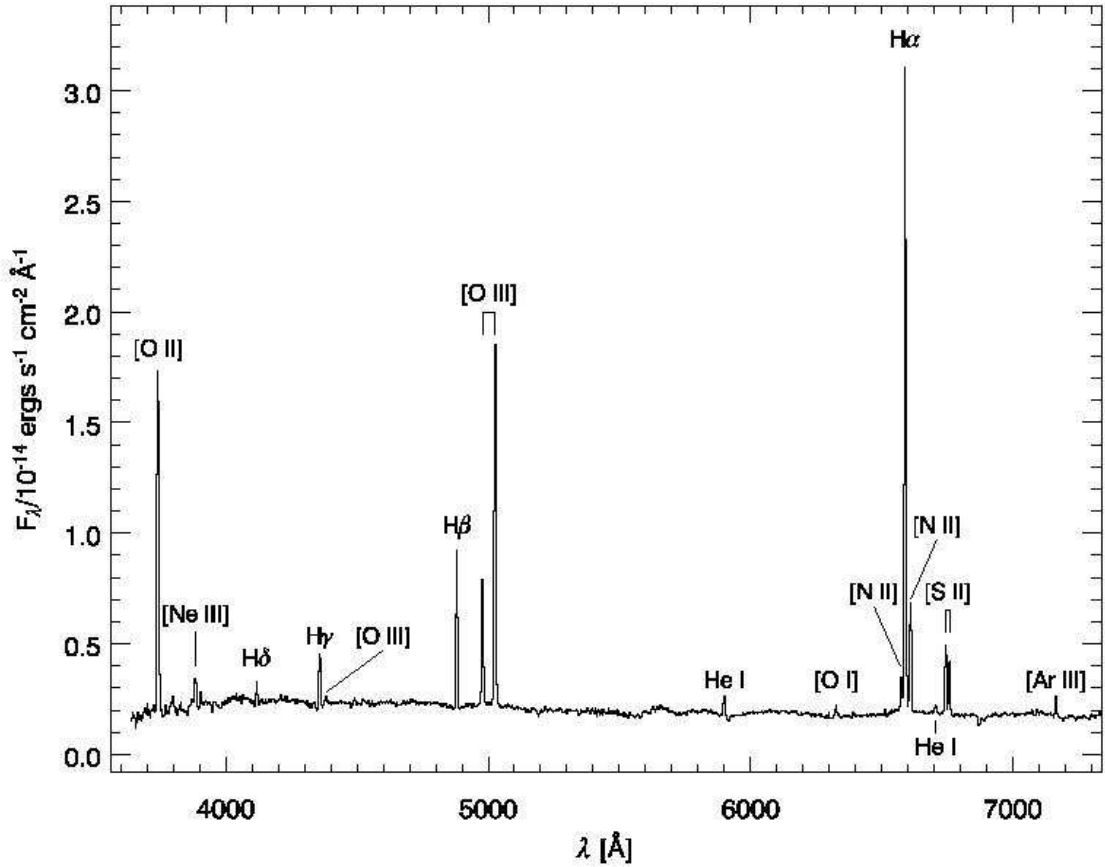}
\end{center}
\caption{Optical spectrum of \ic\ taken with the ARC 3.5-m telescope at Apache Point Observatory. This spectrum was acquired through a $1\farcs5$ slit rotated to a position angle of $90\degr$ and was extracted using a 12\arcsec\ wide aperture. All of the emission lines listed in Table~\ref{tab:eqw} are labelled.
\label{fig:galspec}}
\end{figure}

The inset to Figure~\ref{fig:Halpha} is a pure \Ha\ image of \ic\ and shows ongoing star formation in the nucleus of the galaxy. The stellar continuum was subtracted from our \Ha\ on-band image using a slightly bluer narrowband image. The resulting emission-line image of \ic\ contains flux not only from \Ha, but also from the nearby [\NII]~6548~\AA\ line (the stronger [\NII]~6584~\AA\ line is redshifted out of our \Ha\ on-band filter). We have subtracted the [\NII] flux from our emission-line image by assuming that the entire galaxy has a constant F(\Ha)/F([\NII]~6548~\AA) ratio of 12, as found in our optical spectrum of \ic\ (see \S\,\ref{obs:galspec} and Fig.~\ref{fig:galspec}). This pure \Ha\ image yields a total \Ha\ flux (corrected for Galactic extinction but not intrinsic extinction from \ic) of $(2.9\pm0.1)\times10^{-13}$~ergs~\persec~\persqcm\ for \ic, which corresponds to a \Ha\ luminosity of $(1.03\pm0.04)\times10^{40}\hsq$~ergs~\persec\ at the assumed distance to \ic\ of 17.2\h~Mpc. Using the conversion of \citet{kennicutt98}, this \Ha\ luminosity indicates a current SFR of $0.08\pm0.01\hsq$~\Msun~\peryr\ for \ic. After using the Balmer decrement in our optical spectrum of \ic\ (see \S\,\ref{obs:galspec}, Fig.~\ref{fig:galspec}, Table~\ref{tab:eqw}) to correct for the intrinsic extinction in \ic, we find that its SFR is $0.24\pm0.03\hsq$~\Msun~\peryr. We treat this value as an upper limit on the SFR because we have most likely over-estimated the intrinsic extinction in \ic\ by applying an extinction correction derived from the densest region of the galaxy (i.e., the nuclear region where the Balmer decrement was measured) to our entire \Ha\ image. Thus, we have bounded the SFR of \ic\ to lie in the range 0.08--0.24\hsq~\Msun~\peryr, which is comparable to the values of 0.1--1~\Msun~\peryr\ typically found in nearby dwarf starburst galaxies \citep{martin03}.

\begin{figure}[!ht]
\begin{center}
\epsscale{1.00}
\plotone{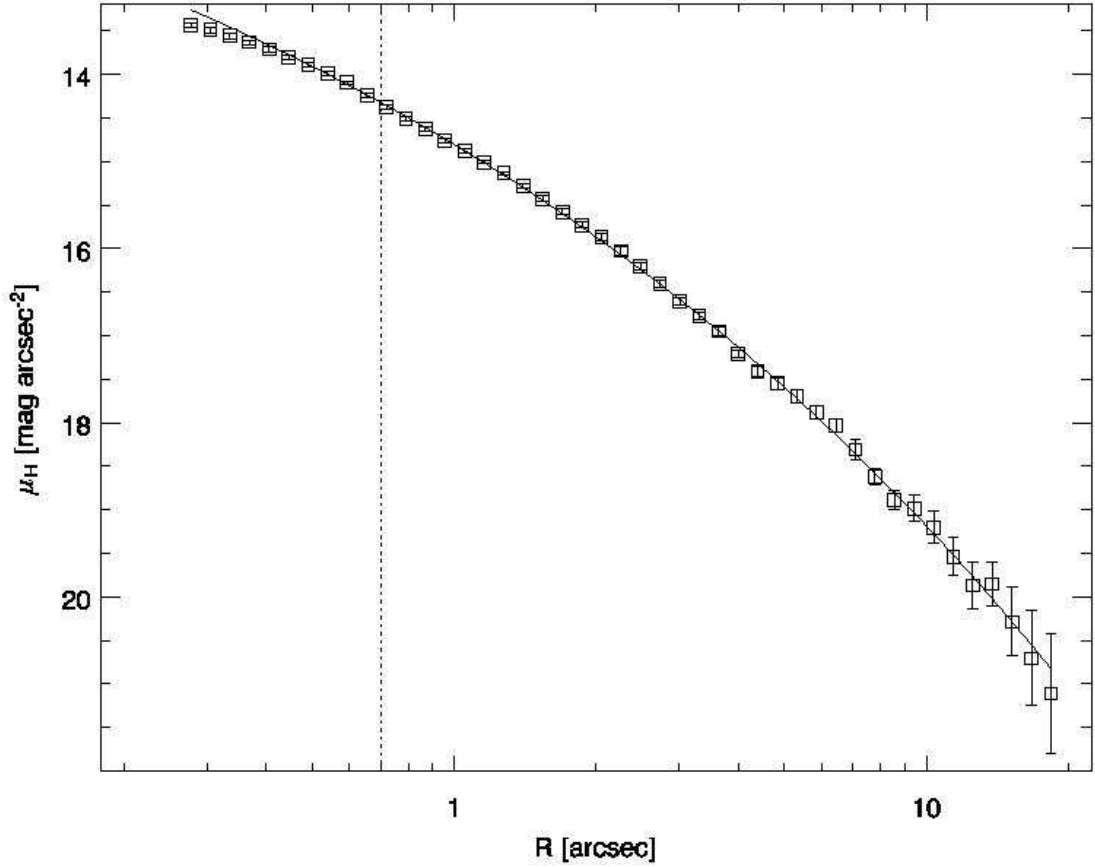}
\end{center}
\caption{Surface brightness profile of \ic\ in H-band. The best-fit (reduced $\chi^2 = 0.6$ for $R \geq 0\farcs7$) \citet{devaucouleurs48} $R^{1/4}$ profile, which has an effective radius of $r_{\rm e} = 400\pm10$\h~pc and a fiducial surface brightness of $\mu_H(r_{\rm e}) = 17.49\pm0.07$~mag~\persqarcsec, is overlaid with a solid line. The seeing of $0\farcs7$ is indicated by a vertical dotted line.
\label{fig:sbprof}}
\end{figure}

Near-infrared H- and K-band images of \ic\ were obtained with the Near Infrared Camera (NIC) at the Apache Point Observatory ARC 3.5-m telescope on 2005 Apr 20. \ic\ was observed for a total exposure time of 100~s in both H- and K-band. We determined the surface brightness profile of the H-band images to find the photometric properties of the stellar distribution of \ic\ since the other broadband images were contaminated by nebular emission lines associated with star formation ([\OII] in B-band, \Ha\ in R-band, and Br$\gamma$ in K-band). Figure~\ref{fig:sbprof} shows the H-band surface brightness profile of \ic, generated with the IRAF tasks {\tt ellipse} and {\tt bmodel}, with a best-fit \citet{devaucouleurs48} $R^{1/4}$ profile overlaid. The dotted vertical line indicates the seeing of $0\farcs7$ in our image. Due to the small field of view of NIC there were no in-field standard stars, so the photometric zero point of $C_H = 23.93\pm0.03$~mag was determined by comparing the 2MASS H-band magnitudes of \ic\ at several apertures to the instrumental isophotal magnitudes at the same apertures. With this calibration, \ic\ has an H-band magnitude, integrated to the radius of the 20th mag~\persqarcsec\ K-band isophote ($r_{K20} = 13\arcsec = 1.1\h$~kpc), of $m_{H,K20} = 11.36\pm0.03$ and the best-fit $R^{1/4}$ profile has an effective radius of $r_{\rm e} = 400\pm10$\h~pc and a fiducial surface brightness of $\mu_H(r_{\rm e}) = 17.49\pm0.07$~mag~\persqarcsec. The ellipticity of the isophotes varies smoothly from $\epsilon=0$ in the center of the galaxy to $\epsilon = 0.3$ at a semi-major axis of $1\farcs1$ ($\sim 90$\h~pc) and stabilizes at this value for larger radii. Thus, \ic\ has an observed axial ratio of 0.7 in H-band\footnote{The SDSS images of \ic\ show that it has an axial ratio of $\sim 0.6$ in $g'$-, $r'$-, and $i'$-bands, which is consistent with the value implied by our R-band image. We adopt the H-band axial ratio of 0.7 because we believe that properties derived from the H-band images are more indicative of the general stellar population underlying the current starburst.}, implying that it is at an inclination of $i = 54\degr\pm2\degr$ using the distribution of intrinsic dwarf galaxy shapes derived by \citet*{staveley-smith92}. Subtracting the model galaxy from our original H-band image reveals a residual stream of material to the north that we interpret as a tidal tail caused by an interaction with a nearby galaxy (see \S\,\ref{obs:galemmap}).

\subsection{Optical Spectrum of \ic}
\label{obs:galspec}

An optical spectrum of \ic\ that covers 3600--9700~\AA\ at $\sim 6$~\AA\ resolution was obtained on 2003 May 2 with the Dual Imaging Spectrograph (DIS) at the Apache Point Observatory 3.5-m telescope. The spectrum is split onto two CCDs with reduced sensitivity from 5200--5600~\AA\ due to a dichroic. The galaxy spectrum was acquired through a $1\farcs5$ slit rotated to a position angle of $90\degr$ and was extracted using a 12\arcsec\ wide aperture. 

Table~\ref{tab:eqw} lists the rest-frame equivalent widths for all emission lines detected in our optical spectrum of \ic. This spectrum is displayed in Figure~\ref{fig:galspec} with all emission lines from Table~\ref{tab:eqw} labelled. The total \Ha\ flux in the spectrum is $2.3\times10^{-13}$~ergs~\persec~\persqcm, or $\sim 80$\% of the flux in the \Ha\ image (\S\,\ref{obs:galimg}). The emission lines in Table~\ref{tab:eqw} are observed at an average velocity of $\langle cz \rangle = 1200\pm30$~\kms, which agrees with the \HI\ 21~cm velocity of $\czgal = 1204\pm3$~\kms\ found in \S\,\ref{obs:galemmap}. We adopt the latter value as the best redshift of \ic\ due to its smaller error bars.

\subsection{\HI\ 21~cm Emission Map of \ic}
\label{obs:galemmap}

\ic\ was observed in the Very Large Array (VLA) D-array on 2003 Apr 23 with a bandwidth of 3.125 MHz centered at 1.422~GHz ($cz = 1200$~\kms) and a channel width of 48.8~kHz (10.4~\kms). The data were reduced using standard AIPS procedures: 3C~286 was used as a flux calibrator while 1035+564 (J2000) was used for the bandpass calibration. The data were then imaged using uniform weighting with the parameter ROBUST=0. After imaging, the data were continuum subtracted using IMLIN. In order to mask the noise in the cube before moment maps were constructed, the cube was first convolved with a Gaussian to a resolution of $100\arcsec \times 100\arcsec$. This lower resolution cube was then blanked at the $2\sigma$ level (1.8~mJy/beam) followed by blanking of the cube by hand to remove features that are not correlated from one channel to the next. This smoothed, blanked cube was then applied as a mask to the original data. The resultant cube was then used to create moment maps and spectra.

Figure~\ref{fig:HI} shows \HI\ 21~cm intensity contours overlaid on a SDSS $g'$-band image of \ic, which indicate that it is interacting with a low surface brightness (LSB) galaxy to the north, SDSS~J112625.96+591737.5 (hereafter \sdsslsb). The SDSS spectrum of the LSB galaxy shows no emission lines and places it at a redshift of $cz = 1340\pm50$~\kms. Interestingly, if we separate the \HI\ 21~cm emission from \ic\ and \sdsslsb\ by placing a cut at a declination of $59\degr14\arcmin$ we find a much lower redshift for \sdsslsb\ of $cz = 1180\pm20$~\kms. This same cut indicates that the \HI\ 21~cm emission from \sdsslsb\ has a FWHM of $26\pm6$~\kms\ and a \HI\ mass of $\MHI = (4.6\pm0.5)\times10^7$~\Msun\ and the \HI\ around \ic\ has a velocity of $\czgal = 1204\pm3$~\kms, a FWHM of $116\pm3$~\kms, and an \HI\ mass of $\MHI = (3.6\pm0.1)\times10^8$~\Msun. 

A tilted ring model was fit to the \HI\ 21~cm velocity profile of \ic, with the ring inclination fixed at the value ($i=54\degr$) predicted by the axial ratio of the optical and near-infrared images (\S\,\ref{obs:galimg}). The approaching (northern) and receding (southern) sides of the galaxy were fit separately to search for evidence of tidal effects and agree to within errors for radii $< 14.2\h$~kpc (170\arcsec). At larger radii the two sides of the galaxy become more and more discrepant, indicating that tidal effects are becoming increasingly important. Our B-band images of \ic\ (\S\,\ref{obs:galimg}) indicate that its Holmberg radius is $66\arcsec = 5.5$\h~kpc, so our 14.2\h~kpc cutoff corresponds to a radius of $\sim 2.5$ Holmberg radii. Figure~\ref{fig:rotcur} shows our final rotation curve for \ic, which was fit to both the approaching and receding sides of the galaxy simultaneously with a fixed inclination of $i=54\degr$ and is truncated at a radius of 14.2\h~kpc. This rotation curve is used to derive the dynamical mass of \ic\ in \S\,\ref{esc:mass}.

\section{Galaxy/Absorber Connection}
\label{galabs}

Statistically, low column density ($\NHI \lesssim 10^{17.3}$~\persqcm) intergalactic \Lya\ absorbers are associated with overdense IGM filaments rather than individual luminous ($L > 0.1$~\Lstar) galaxies \citep*{morris93,tripp98,impey99,penton02,penton04}. Existing HST and FUSE spectra of QSO sight lines with low redshift intergalactic \Lya\ absorbers are not generally sensitive enough to detect metals (\CII, \CIII, \CIV, \SiII, \SiIII, \SiIV, \OVI) in these absorbers except at $\NHI \gtrsim 10^{14}$~\persqcm\ \citep*{stocke06a} where metallicities of 5--10\% solar are found \citep*[but see \citealt{aracil06} and \citealt{prochaska04} for individual absorber metallicities that may approach solar values]{sembach01,tripp02,shull98,shull03,tumlinson05b,danforth05,danforth06}. Weak metal-line systems are found to have smaller impact parameters from nearby galaxies than typical \Lya\ absorbers without associated metal lines \citep[e.g.,][]{stocke06b} and the average QSO/galaxy impact parameter decreases with increasing column density. Thus, Damped \Lya\ Absorbers (DLAs), the highest column density systems ($\NHI > 10^{20.3}$~\persqcm), are plausibly associated with galactic disks (although see \citealp{york06} for a different hypothesis) and Lyman limit systems (LLSs; $\NHI = 10^{17.3-20.3}$~\persqcm) are likely associated with galactic halos \citep[e.g.,][]{steidel95,steidel98}, whereas lower column density metal-line absorbers ($\NHI = 10^{14-17.3}$~\persqcm) could be associated with outflowing galactic winds \citep[see, e.g.,][]{stocke04,stocke06a,stocke06b}. These statistical results suggest that the \Lya+\CIV\ absorber at $\czabs = 1110\pm30$~\kms\ in the spectrum of \sbs\ could be associated with a nearby galaxy. 

\begin{figure}[!ht]
\begin{center}
\epsscale{0.95}
\plotone{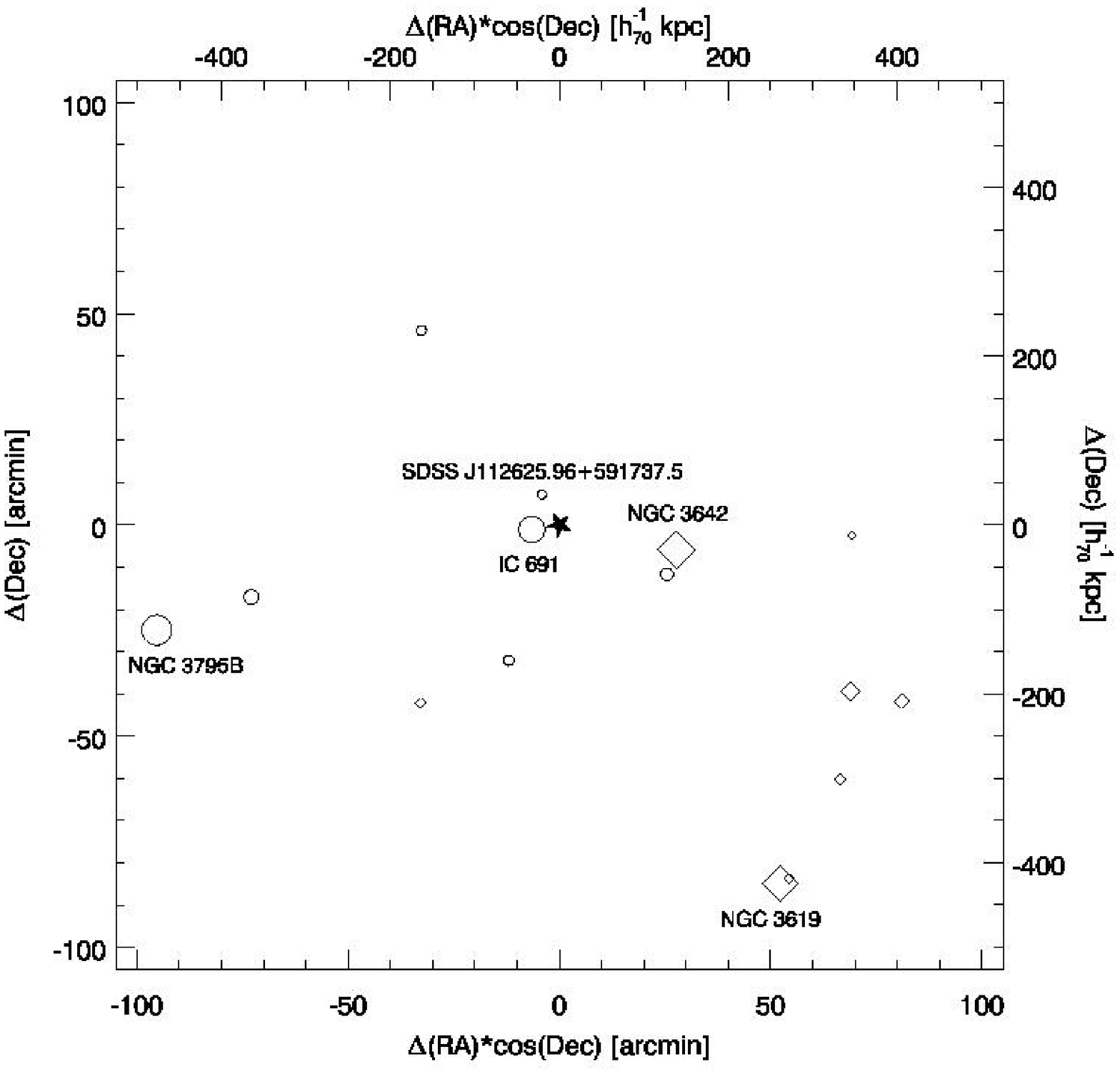}
\end{center}
\caption{All galaxies with $cz < 1600$~\kms\ that are within 500\h~kpc of \sbs\  ($100\arcmin$ at $cz = 1200$~\kms), as compiled by SDSS and NED. All positions are relative to \sbs\ (${\rm RA} = 11^{\rm h}\,25^{\rm m}\,53.8^{\rm s}$, ${\rm Dec} = 59\degr\,10\arcmin\,22\arcsec$), which is represented by the filled star symbol in the center of the plot. Circles represent galaxies with $cz < 1300$~\kms\ and diamonds represent galaxies with $cz = 1300$--1600~\kms. Symbol size is proportional to SDSS $r'$-band luminosity, with a factor of 2 increase in symbol size equivalent to a factor of 10 increase in luminosity. The galaxies discussed in \S\,\ref{galabs} are also labelled.
\label{fig:galenv}}
\end{figure}

Figure~\ref{fig:galenv} shows the positions of all galaxies with $cz < 1600$~\kms\ that are within 500\h~kpc of \sbs\ ($100\arcmin$ at $cz = 1200$~\kms), as compiled by SDSS and the NASA/IPAC Extragalactic Database (NED). The position of \sbs\ is indicated by the filled star symbol in the center of the plot. The circular symbols represent galaxies with $cz < 1300$~\kms\ and the diamonds represent galaxies with $cz = 1300$--1600~\kms. Symbol size is proportional to the SDSS $r'$-band luminosity of the galaxy. While no obvious groups of galaxies are evident, there is a large-scale ``filament'' of galaxies present in Figure~\ref{fig:galenv} in which both \ic\ and the absorber are embedded. This absorber environment is similar to the filamentary environments of the 1586~\kms\ absorber towards 3C~273 and the 1685~\kms\ absorber towards RX~J1230.8+0115 studied previously by \citet{stocke04} and \citet{rosenberg03}. Any galaxies with unknown redshifts near \sbs\ are fainter than the SDSS spectroscopic completeness limit of $m_r = 17.8$. At $\czgal \approx \czabs$, this limit implies that Figure~\ref{fig:galenv} is complete to an absolute magnitude of $M_r - 5\log{h_{70}} \leq -13.2$, or $L_r \geq 0.001\hsq$~\Lstar\ \citep{blanton01}.

With impact parameters of 33\h~kpc and 42\h~kpc, respectively, \ic\ and \sdsslsb\ are by far the closest galaxies in Figure~\ref{fig:galenv} to the \sbs\ sight line. While the redshift of \sdsslsb\ as derived from the centroid of its \HI\ 21~cm emission is marginally closer to the absorber velocity than that of \ic, \ic\ is $\sim 20$ times more luminous in SDSS $r'$-band than \sdsslsb. This luminosity difference, combined with the facts that \ic\ is currently forming stars while \sdsslsb\ is not and \ic\ is closer to the \sbs\ sight line, indicate that \ic\ is more likely to have created the metal-line absorber.

The closest luminous galaxy to \sbs\ is NGC~3642 at $\czgal = 1598$~\kms, which is 130\h~kpc away ($28\farcm3$ at \czabs) and $\sim 10$ times more luminous than \ic. We consider \ic\ to be the more likely source of the metal-line absorber, however, since NGC~3642 is both 4 times further from \sbs\ in projection and 5 times more discrepant in redshift. Other than NGC~3642, the only galaxies in Figure~\ref{fig:galenv} that are more luminous than \ic\ are NGC~3795B and NGC~3619, which are 452\h~kpc ($97\farcm9$ at \czabs) and 458\h~kpc ($99\farcm2$ at \czabs) away from \sbs, respectively. NGC~3795B is at a redshift of $cz = 1257$~\kms\ and is twice as luminous as \ic, and NGC~3619 is at a redshift of $cz = 1553$~\kms\ and is $\sim 6$ times more luminous than \ic. Neither of these galaxies are as luminous as NGC~3642 nor as close to the \sbs\ sight line so they are not likely candidates for the origin of the metal-line absorber.

\begin{figure}[!ht]
\begin{center}
\epsscale{0.95}
\plotone{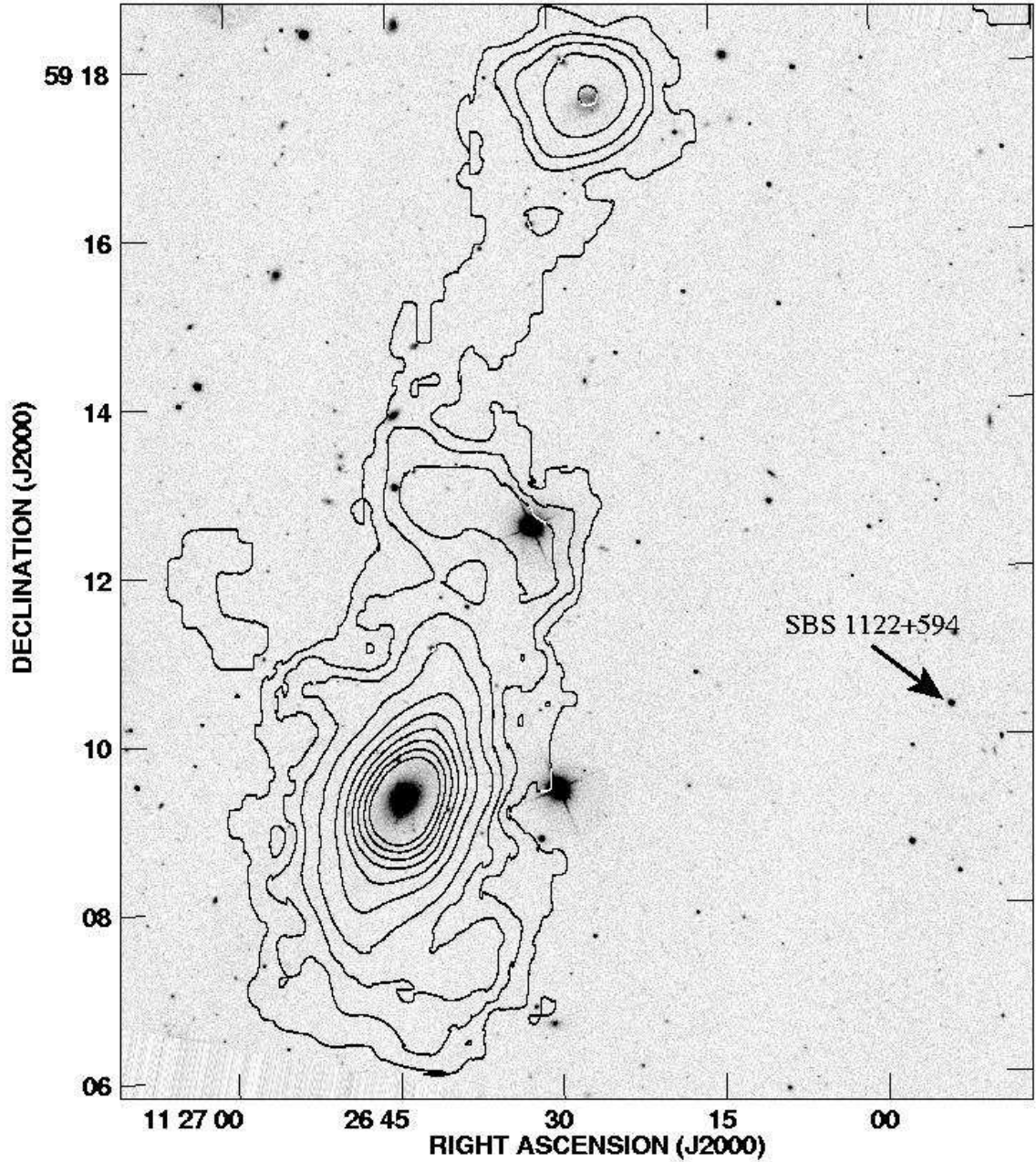}
\end{center}
\caption{\HI\ 21~cm intensity contours for \ic\ overlaid on a SDSS $g'$-band image. Contour levels are at 1.78, 8.92, 17.8, 35.7, 71.4, 107, 143, 178, 214, and $250 \times 10^{18}$~atoms~\persqcm. \ic\ is interacting with \sdsslsb, an LSB galaxy to the north. The position of \sbs\ is also labelled.
\label{fig:HI}}
\end{figure}

Several circumstantial lines of reasoning also point to a connection between \ic\ and the metal-line absorber. The 33\h~kpc impact parameter between \sbs\ and \ic\ is much less than the median nearest-neighbor distance of 180\h~kpc between low-metallicity ($10\pm5$\% solar) \OVI\ and \CIII\ absorbers detected with FUSE and galaxies of any luminosity detected in regions where galaxy surveys are complete to at least 0.1~\Lstar\ \citep{stocke06b}. The \sbs\ sight line also lies at a position angle $\theta = 27\degr$ from the minor axis of \ic, which is the expected direction for an outflowing starburst wind since nearby starburst galaxies often show biconic outflows with opening angles above the disk of $2\theta \approx 45\degr$--$100\degr$ \citep[values of $2\theta \sim 65\degr$ are typical;][]{heckman90,veilleux05}. Furthermore, the \HI\ 21~cm emission along the major axis of \ic\ extends to distances comparable to the \sbs/\ic\ separation (see Fig.~\ref{fig:HI}). While some of this material is affected by tidal interactions between \ic\ and \sdsslsb, the large extent of the \HI\ envelope of \ic\ with respect to the angular distance between \ic\ and \sbs\ indicates that gas associated with \ic\ can plausibly reach the location of \sbs\ on the sky.

The \HI\ emission map in Figure~\ref{fig:HI} suggests that the metal-line absorber could be due to the recent interaction between \ic\ and \sdsslsb. However, the extended \HI\ is in a plane nearly perpendicular to the direction of \sbs\ from \ic, and while tidal tails are often strongly curved \citep[e.g., NGC~520;][]{hibbard96,norman96} the \HI\ 21~cm emission of \ic\ shows no evidence of such curvature. Also, the \HI\ velocities of the tidal border between these two galaxies is never less than $1180\pm20$~\kms\ at any location. Therefore, any tidal debris in the direction of \sbs\ must be more diffuse and/or more highly-ionized and have different kinematics than the tidal material seen in Figure~\ref{fig:HI}. The absence of velocity overlap between the \HI\ tidal debris and the absorber argue against a tidal origin for the absorbing gas. So, while we cannot rule out tidal debris as the origin of the metal-line absorber, we believe that the outflowing gas from a starburst superwind triggered by the recent interaction is more likely. This model naturally explains why absorption is seen near the minor axis of \ic, which is far from the plane of the galaxy interaction \citep*[although outflows from dwarf galaxies may have less of a preferred direction than in more massive starbursts; e.g.,][]{ott03}.

Table~\ref{tab:eqw} lists the rest-frame equivalent widths for the emission lines detected in the \ic\ starburst with our optical spectrum (\S\,\ref{obs:galspec}). The [\OIII] derived mean temperature of $\sim14,000\degr$~K is uncertain due to the weakness of the [\OIII]~4363~\AA\ line in our spectrum, but is similar to [\OIII] derived temperatures for hotter \HII\ regions in the LMC \citep{oey00a}. Likewise the relatively reddening-free line ratios [\OIII~5007\,\AA]/\Hb\ and [\NII~6584\,\AA]/\Ha\ in \ic\ are also similar to values found for LMC \HII\ regions by \citet{oey00b}. Because the Balmer decrement listed in Table~\ref{tab:eqw} is only slightly larger than case-B, there is only quite modest reddening present in the \HII\ region spectrum of \ic\ and should not effect these line ratios significantly. Since the detailed position-resolved abundance study of LMC \HII\ regions by \citet{oey00a} found oxygen and nitrogen abundances of 20--30\% solar, the mean metal abundances in the \ic\ starburst are comparable to those values based upon our long-slit spectrum. Therefore, the metal abundance in this dwarf starburst is significantly subsolar as expected for such a small galaxy.

Our STIS spectrum of \sbs\ (see \S\,\ref{obs:qsospec} and Fig.~\ref{fig:qsospec}) limits the \HI\ and \CIV\ column densities to $\NHI \geq 10^{14.6}$~\persqcm\ and $\NCIV = 10^{14.7\pm0.2}$~\persqcm, respectively. A standard photoionization model can be employed to estimate the metallicity of this absorber by assuming an extragalactic photoionizing flux of $\Inu = 10^{-23}$~ergs~\persqcm~\persec~\perHz\ and a gas overdensity for this absorber of 10--100 based upon estimates from numerical simulations for the observed $\NHI$ lower limit (i.e., $\log{\rm U} = -0.5$ to $-1.5$). In this case the \CIV\ equivalent width of 800~m\AA\ yields  $[Z/\Zsun] < -0.3$ and a particle density of $\sim 10^{-5}$~\percubcm, implying a scale-size along the line-of-sight (LOS) of $>10$~pc. It is quite unlikely that this absorber is either hotter, collisionally-ionized gas or diffuse photoionized gas because of the strength of the \CIV\ absorption; however, searches for \OVI\ in the far-UV and lower ions in the HST band should be conducted if better UV spectra become available to discriminate between these two possiblities. Our estimated absorber metallicity is broadly consistent with our metallicity estimate of $[Z/\Zsun] \sim -0.7$ for \ic, although the absorber metallicity could be less than that of \ic\ since there is only a lower limit on the absorber \NHI. However, the upper limit on the absorber metallicity is still significantly subsolar as expected from an outflow associated with a dwarf galaxy.

While the inferred LOS size for this absorber may seem quite small, it is comparable to the LOS scale-size for the 1586~\kms\ absorber in the 3C~273 sightline \citep{tripp02}, as well as many ``weak-\MgII'' absorbers studied by \citet{charlton02} at higher redshifts. \citet{stocke04} argued that the small LOS sizes reported for weak metal-line absorbers coupled with their relative frequency (which requires a scale-size on the sky of $\sim 100$~kpc) were naturally explained by the thin-shell geometry produced by a galactic superwind. Recent hydrodynamical simulations of dense clouds entrained in an outflowing galactic superwind by \citet{marcolini05} modelled these clouds as spheres with a radius of 15~pc, which is comparable to the LOS sizes of the 3C~273 and \sbs\ absorbers discussed above. Therefore, both the subsolar metallicity of the metal-line absorber and its small size are suggestive of an origin in a dwarf galaxy superwind from \ic.

\section{Does the Starburst Wind Escape?}
\label{escape}

In this Section, we assume that the \Lya+\CIV\ absorber in \sbs\ is gas entrained in a starburst superwind from \ic. Since galactic winds are a leading mechanism for transporting metals and energy from galaxies into the IGM, we use the \Lya+\CIV\ absorber in \sbs\ to determine whether the starburst wind from \ic\ can escape from its gravitational potential to enrich large regions of intergalactic space. The total mass of \ic\ is calculated in \S\,\ref{esc:mass}. In \S\,\ref{esc:wind} we use this mass to derive various properties of the starburst wind.

\subsection{Total Mass of \ic}
\label{esc:mass}

\begin{figure}[!ht]
\begin{center}
\epsscale{1.00}
\plotone{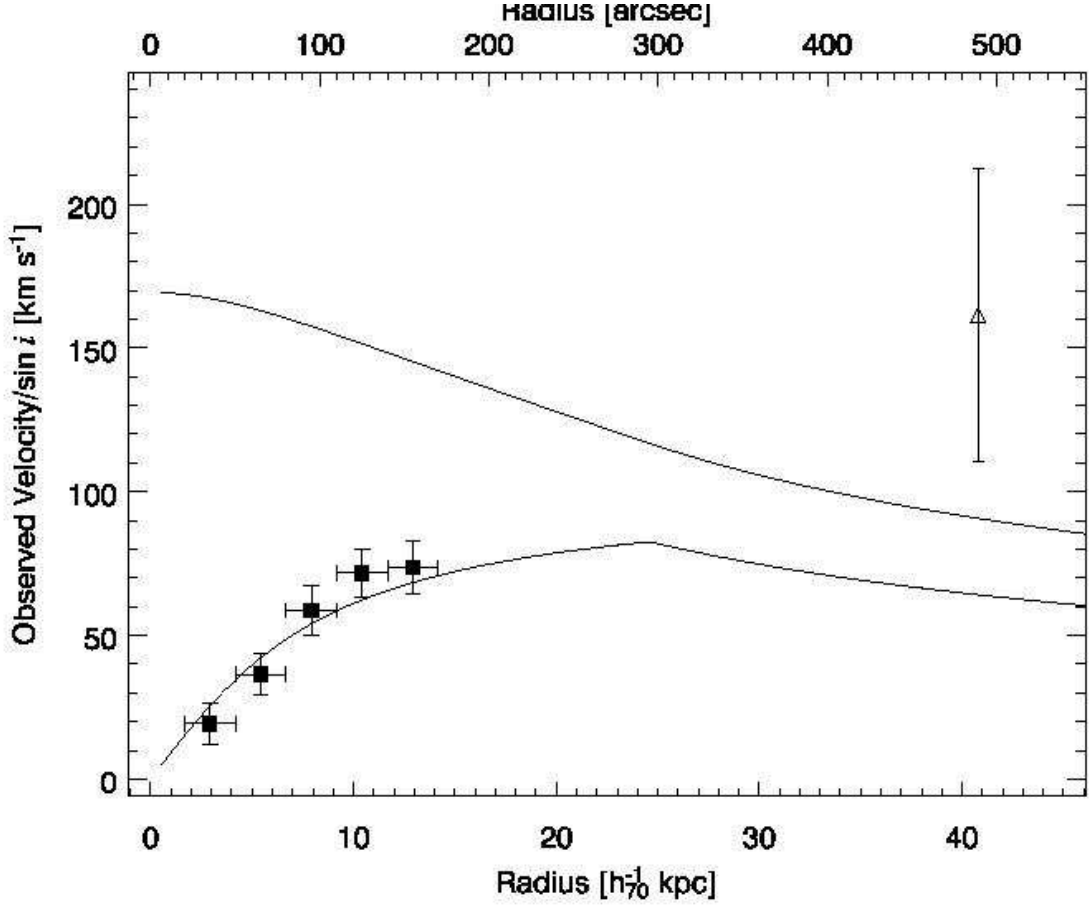}
\end{center}
\caption{Rotation curve of \ic\ with a truncated isothermal halo model overlaid. The model assumes that the rotation curve is flat to a radius of 24.6\h~kpc (295\arcsec) and has a best-fit central density of $\rho_0 = 0.005\pm0.002$~\Msun~\percubpc\ and a core radius of $6\pm2$\h~kpc. The solid line above the data points indicates the escape velocity as a function of radius predicted by the best-fit model. The data point with the open triangle symbol shows the derived wind velocity and absorber location (\S\,\ref{esc:wind}). All quantities assume that \ic\ is at an inclination of $i=54\degr$.
\label{fig:rotcur}}
\end{figure}

The \HI\ 21~cm rotation curve discussed in \S\,\ref{obs:galemmap} is shown in Figure~\ref{fig:rotcur}. These data are truncated at a radius of 14.2\h~kpc (170\arcsec) because the interaction of \ic\ with \sdsslsb\ (the LSB galaxy to the north; see Fig.~\ref{fig:HI}) precludes us from extending the rotation curve to larger radii. However, the last data point in Figure~\ref{fig:rotcur} suggests that the \HI\ 21~cm rotation curve has begun to flatten at radii $>9.2\h$~kpc (110\arcsec). In order to estimate the total mass of \ic, we have assumed that the rotation curve has flattened at these radii and remains flat to a radius of 24.6\h~kpc (295\arcsec; $\sim 4.5$ Holmberg radii), the maximum radial extent of \ic\ (using the cut at a declination of $59\degr14\arcmin$ to separate the \HI\ emission from \ic\ and \sdsslsb; see \S\,\ref{obs:galemmap}). Under this assumption, the rotation curve was fit with an isothermal halo truncated at 24.6\h~kpc. The best-fit model (reduced $\chi^2 = 1.0$) is overlaid on the data points in Figure~\ref{fig:rotcur} and has a central density of $0.005\pm0.002$~\Msun~\percubpc\ and a core radius of $6\pm2\h$~kpc. The escape velocity as a function of radius predicted by the best-fit model is shown in Figure~\ref{fig:rotcur} by the solid line above the data points, and the point with the open triangle symbol indicates the observed wind velocity and absorber location. These quantities are discussed further in \S\,\ref{esc:wind}.

Assuming that the rotation curve of \ic\ is flat at a velocity of $74\pm9$~\kms\ to a radius of 24.6\h~kpc and declines thereafter implies that \ic\ has a total mass of $\Mdyn = (3.1\pm0.5)\times10^{10}\h$~\Msun. This dynamical mass predicts a total mass-to-light ratio of $\Mdyn/\LB = 28\pm5\,h_{70}$ and a gas mass fraction of $f_{\rm gas} \equiv \MHI/\Mdyn = 0.012\pm0.002\,h_{70}$ for \ic. These values are consistent with the mass-to-light ratios and gas mass fractions found for nearby blue compact galaxies and dwarf irregular galaxies \citep{begum05,pisano01,roberts94}.

It is quite possible that this dynamical mass is an overestimate. Figure~\ref{fig:HI} shows that \ic\ is interacting with a nearby galaxy to the north, which should act to increase the turbulent motions in \ic, and thus its velocity dispersion. Therefore, one would expect that only part of the velocity in Figure~\ref{fig:rotcur} is caused by galactic rotation with the remainder caused by the interaction. If this is correct it implies that the true dynamical mass of \ic\ is lower than our estimate. \ic\ is also well-fit by a \citet{devaucouleurs48} $R^{1/4}$ profile (Fig.~\ref{fig:sbprof}), making it photometrically similar to elliptical galaxies rather than spirals, further implying that it may not be dominated by rotation. However, since our dynamical mass estimate of $\Mdyn = (3.1\pm0.5)\times10^{10}\h$~\Msun\ predicts a mass-to-light ratio and gas mass fraction for \ic\ that are consistent with the range of values found for similar nearby galaxies, we will assume that it is valid for all subsequent calculations.

\subsection{Starburst Wind Properties}
\label{esc:wind}

We assume that the \Lya+\CIV\ absorber is entrained in a radial outflow emanating from the galactic center of \ic, that the galaxy's mass distribution is spherically symmetric, and that all of the mass is located interior to the \sbs\ sight line. Under these assumptions, \ic\ can be treated as a point-mass when calculating the escape velocity at the absorber location: 
\begin{equation}
\vesc = 93~\kms \left(\frac{\Mdyn}{10^9~\Msun}\right)^{1/2} \left(\frac{r}{\rm kpc}\right)^{-1/2}
\end{equation}
If \ic\ is at an inclination $i$ then $\Mdyn = ([2.0\pm0.3]\times10^{10}\h~\Msun)/\sin^2{i}$ (which corresponds to the value from \S\ref{esc:mass} for $i=54\degr$), $r = (33\h~{\rm kpc})/\sin{i}$, and $\vesc = (72\pm8~\kms)/\sqrt{\sin{i}}$. The starburst wind has a LOS velocity of $\vlos \equiv |\czabs - \czgal| = 95\pm30$~\kms\ at the absorber location, which corresponds to an outflow speed perpendicular to the disk of \ic\ of $\vw = \vlos/\cos{i}$. The wind will escape the gravitational potential of \ic\ if $\vw/\vesc > 1$, which will occur at inclinations of $i > 26\degr\pm10\degr$. The H-band axial ratio (\S\,\ref{obs:galimg}) predicts an inclination of $i=54\degr\pm2\degr$ for \ic. Thus, under our set of assumptions, \ic\ produces an unbound starburst wind.

With our assumed geometry and a galaxy inclination of $i=54\degr$, the escape velocity at the absorber location of $r = (33\h~kpc)/\sin{i} = 41\h$~kpc is $\vesc = (72\pm8~\kms)/\sqrt{\sin{i}} = 80\pm10$~\kms\ and the starburst wind is moving at a speed of $\vw = \vlos/\cos{i} = 160\pm50$~\kms. This outflow velocity places an upper limit on the time since the absorbing gas was ejected of $\tau_{\rm ej} = r/\vw = 250\pm80$\h~Myr, assuming that the wind has been moving at a constant velocity. This timescale is an upper limit since the ejecta would likely decelerate due to gravity and mass loading, and indicates that the metal-line absorber is associated with the current star formation episode in \ic\ or a burst immediately preceding it. 

It is interesting to consider how much mass the current burst in \ic\ could eject as wind material. If the burst that created the metal-line absorber lasted for $10^8$~years at the current SFR of \ic\ (uncorrected for intrinsic extinction), then $\sim 8\times10^6$~\Msun\ of stars were formed. A Salpeter initial mass function predicts that a burst of this size would create $\sim 10^5$~stars with $M > 8$~\Msun, each of which would eventually become a supernova. If each supernova has $10^{51}$~ergs of energy available, which is converted to bulk motions with an efficiency of 3--30\% \citep{cioffi91,koo92a,koo92b}, then the burst produced $3\times10^{54-55}$~ergs that could be used to generate a wind. This energy can accelerate up to $6\times10^{6-7}$~\Msun\ of material to a velocity of $\sim 160$~\kms\ after escaping the gravitational potential well of \ic. Thus, the production of the metal-line absorber towards \sbs\ by the \ic\ starburst is plausible energetically.

If \ic\ continues to form stars at its current rate of $0.08\pm0.01\hsq$~\Msun~\peryr\ then it will run out of available material for additional star formation in the gas depletion timescale of $\tau_{\rm gas} \equiv \MHI/{\rm SFR} = 4.5\pm0.6\,h_{70}^2$~Gyr \citep{kennicutt83}. This timescale is longer than the duration of a typical star formation episode, so \ic\ will likely experience further star formation episodes once the current burst ceases. Complicated, episodic star formation histories are not uncommon for dwarf starbursts, as evidenced by the Local Group dwarf elliptical and spheroidal galaxies \citep{mateo98}.

\section{Conclusions}
\label{conclusion}

We have used HST to detect \Lya\ and \CIV\ absorption ($\czabs = 1110\pm30$~\kms) from the nearby dwarf starburst galaxy \ic\ ($\czgal = 1204\pm3$~\kms) in the spectrum of the QSO \sbs. Narrowband \Ha\ images show that \ic\ is currently forming stars at a rate of 0.08--0.24\hsq~\Msun~\peryr, which is comparable to the SFRs found in other nearby dwarf starburst galaxies \citep{martin03}. A long-slit optical spectrum of \ic\ reveals \Hb, H$\gamma$, [\OII], [\OIII], [\NII], and [\SII] emission in addition to \Ha. The H-band surface brightness profile of \ic\ is well-fit by a \citet{devaucouleurs48} $R^{1/4}$ profile, making it photometrically similar to elliptical galaxies despite its current starburst. An \HI\ 21~cm emission map obtained with the VLA shows that \ic\ is interacting with \sdsslsb, a LSB galaxy to the north (see Fig.~\ref{fig:HI}), which could have triggered its current epoch of star formation. Eliminating as much tidal material as possible by-eye, \ic\ has a \HI\ mass of $\MHI = (3.6\pm0.1)\times10^8$~\Msun\ and its \HI\ 21~cm rotation curve (Fig.~\ref{fig:rotcur}) implies that it has a dynamical mass of $\Mdyn = (3.1\pm0.5)\times10^{10}\h$~\Msun.

Since \ic\ is a dwarf starburst galaxy and the \sbs\ sight line lies in the expected direction of a starburst wind \citep{heckman90,veilleux05}, we suggest that the metal-line absorber at $\czabs = 1110\pm30$~\kms\ is caused by an outflowing starburst wind from \ic. If this is indeed the case, then the wind produced by \ic\ will escape its gravitational potential to enrich the surrounding IGM since the escape velocity at the absorber location is $\vesc = 80\pm10$~\kms\ and the wind velocity is $\vw = 160\pm50$~\kms. If the absorbing gas was ejected from \ic\ when it was forming stars at its current rate in a burst lasting $\sim 10^8$~years, then it could accelerate $\gtrsim 10^7$~\Msun\ of material to the current absorber velocity. A rough estimate of a subsolar metallicity ($[Z/\Zsun] < -0.3$) for this absorber is also consistent with the metallicity limit of $[Z/\Zsun] \sim -0.7$ found for \ic.

Our conclusion that \ic\ produces an unbound starburst wind agrees with the results of \citet{stocke04}, who found that an unbound wind from a dwarf poststarburst galaxy could be responsible for the $\czabs = 1586$~\kms\ metal-line system in the 3C~273 sight line 71\h~kpc away. This dwarf poststarburst galaxy could be representative of a later stage in the evolution of the \sbs/\ic\ system. Once all of the \HI\ in \ic\ has been exhausted via star formation and ejection it will develop a poststarburst spectrum and fade in luminosity as envisioned by \citet{babul92} to the current brightness of the 3C~273 dwarf ($\MB = -13.9$) within $\sim 1$~Gyr after the \HI\ is exhausted \citep{bruzual93,bruzual03,babul96}. Meanwhile, the unbound wind will continue to propagate into the surrounding IGM and increase the distance that \ic\ has distributed metals.

On the other hand, more luminous galaxies appear to produce bound winds. \citet{keeney05} found that the nearby luminous (0.5~\Lstar) starburst galaxy NGC~3067 (${\rm SFR} \approx 1.4$~\Msun~\peryr) produces a bound wind along the 3C~232 sight line, which probes the halo of NGC~3067 near its minor axis and 11\h~kpc from the plane. Similarly, \citet{keeney06} found that the Milky Way produces a bound wind toward two high-latitude AGN sight lines near $l = 0\degr$ (Mrk~1383 and PKS~2005--489) that probe regions directly to the north and south of the Galactic Center at heights of up to 12.5~kpc. In both NGC~3067 and the Milky Way, the bound winds have the same spectral signature as high-velocity clouds and share many of their properties. 

Statistical studies of QSO-absorber pairs in large samples of low-$z$ \Lya+metal line absorbers \citep[e.g.,][]{stocke06b} also suggest that IGM metals are spread primarily by dwarf galaxies. For example, \citet{tumlinson05a} found that enriched regions of gas must extend $\sim 1$~Mpc from \Lstar\ galaxies to be due to $\geq \Lstar$ galaxy superwinds. Enrichment regions of 100--150~kpc are much more plausible based upon observed absorber-galaxy distances \citep{stocke06b}, but require that enrichment is due primarily to dwarf ($\lesssim 0.1$~\Lstar) galaxies \citep{tumlinson05a}. The \sbs/\ic\ system supports this conclusion.

\ic\ is not the only blue compact galaxy that shows evidence for an outflowing wind. The Fornax Cluster galaxy FCC~35 shows strong \Ha\ and [\OIII] emission and an unusual single-dish \HI\ 21~cm profile in which a rotationally supported \HI\ disk is superimposed with an irregularly shaped \HI\ cloud with no optical counterpart \citep{putman98}. The disk and the cloud have roughly the same \HI\ mass ($\MHI = 2.2\times10^8$~\Msun) and overlap spatially but not in velocity, with the cloud blueshifted by $\sim 150$~\kms\ with respect to the systemic velocity of the disk. While the cloud has no velocity structure, the rotation curve of the disk indicates that it has a truncated mass distribution. \citet{putman98} suggest that the \HI\ cloud is triggering the current burst of star formation in FCC~35, but by analogy with \ic\ the \HI\ cloud could also be ejecta from the starburst wind of FCC~35, which would explain the truncated mass distribution of its disk. FCC~35 has an \HI\ mass comparable to the galaxies with the lowest gas masses and smallest gas depletion timescales in the \citet{pisano01} sample of nearby blue compact galaxies. These low-mass galaxies may be undergoing their last star formation event and will likely fade in luminosity as envisioned by \citet{babul92} once their current star formation ceases \citep{pisano01}.

Collectively, these results imply that starburst winds escape more easily from dwarf starburst galaxies than their more massive counterparts. This is to be expected since \citet{meurer97} found a maximum areal SFR of $\sim 45$~\Msun~\persqkpc~\peryr\ for starbursts of all luminosities and \citet{martin99} found that the temperature of starburst winds is nearly constant as a function of galaxy mass, both of which imply that the strength of a starburst wind is independent of galaxy size. Thus, there is growing evidence that the metals and energy expelled by massive galaxies are retained in their bound halos and the weaker but more numerous dwarfs are primarily responsible for enriching the IGM.

We have argued that an outflowing unbound starburst wind from \ic\ is the best model for explaining the origin of the metal-line absorber in the spectrum of \sbs\ due to: (1) the proximity of the QSO sight line to the galaxy's minor axis, (2) the on-going star formation in \ic, (3) the energetic ability of the current \ic\ starburst to eject $\sim 10^7$~\Msun\ of material to the observed wind velocity after escaping the galaxy's gravitational potential well, and (4) the consistent subsolar metallicities derived for \ic\ and the metal-line absorber. However, other plausible models for the absorber origin exist. In particular, we cannot rule out the possibility that the absorber is caused by diffuse tidal debris from the recent interaction of \ic\ and \sdsslsb, although the observed geometry of the \HI\ 21~cm emitting gas does not obviously support this hypothesis. Future observations of \sbs\ as well as fainter AGN near \ic\ with the Cosmic Origins Spectrograph (COS) would allow us to distinguish between these models and sensitively search for absorption in other ions at $cz \sim 1110$~\kms.

{\acknowledgments We acknowledge support from NASA HST General Observer grant GO-09874 to the University of Chicago, and grants GO-09506 and GO-06593 to the University of Colorado. B. A. K. acknowledges support from NASA Graduate Student Researchers Program grant NGT5-154. J. L. R. acknowledges support from NSF grant AST-0302049. J. T. gratefully acknowledges the generous support of the Gilbert and Jaylee Mead Family Foundation through the Yale Department of Physics. We also thank Mark Giroux for use of his photoionization models.

This work is based on observations made with the NASA/ESA {\em Hubble Space Telescope} (HST), the Apache Point Observatory (APO) 3.5-meter telescope, the Kitt Peak National Observatory (KPNO) 2.1-meter telescope, and the Very Large Array (VLA) of the National Radio Astronomy Observatory (NRAO). The HST data were obtained at the Space Telescope Science Institute, which is operated by the Association of Universities for Research in Astronomy, Inc., under NASA contract NAS5-26555. The APO 3.5-m telescope is owned and operated by the Astrophysical Research Consortium. KPNO is operated by the Association of Universities for Research in Astronomy, Inc. (AURA) under cooperative agreement with the National Science Foundation. The NRAO is a facility of the National Science Foundation operated under cooperative agreement by Associated Universities, Inc. This research has also made use of the NASA/IPAC Extragalactic Database (NED) and the Sloan Digital Sky Survey (SDSS). NED is operated by the Jet Propulsion Laboratory, California Institute of Technology, under contract with the National Aeronautics and Space Administration, }

\appendix
\section{Appendix}

The full STIS G140L snapshot spectrum of \sbs\ is displayed in Figure~\ref{fig:fullqsospec}.  This spectrum shows a $\sim 30$\% continuum decrement blueward of $\sim1435$~\AA, which suggests the presence of a partial Lyman limit system (LLS) at $z\sim0.58$. A search for other absorption lines at this redshift revealed Ly$\beta$, \CII/\CII*, and \OI\ absorption at $z=0.583$. The apparent optical depth \citep[AOD;][]{savage91} of the Ly$\beta$ line places a lower limit on the \HI\ column density of the LLS of $\NHI > 10^{15.4}$~\persqcm. LLSs have column densities of $\NHI = 10^{17.3-20.3}$~\persqcm\ and an optical depth at the Lyman limit of $\tau_{\rm LL} \gtrsim 1$. The $28\pm13$\% continuum decrement in our spectrum of \sbs\ implies an optical depth at the Lyman limit of $\tau_{\rm LL} = 0.3\pm0.2$, which requires an \HI\ column density of $\NHI = (5\pm3)\times10^{16}$~\persqcm.

\begin{figure}[!ht]
\begin{center}
\epsscale{0.7}
\plotone{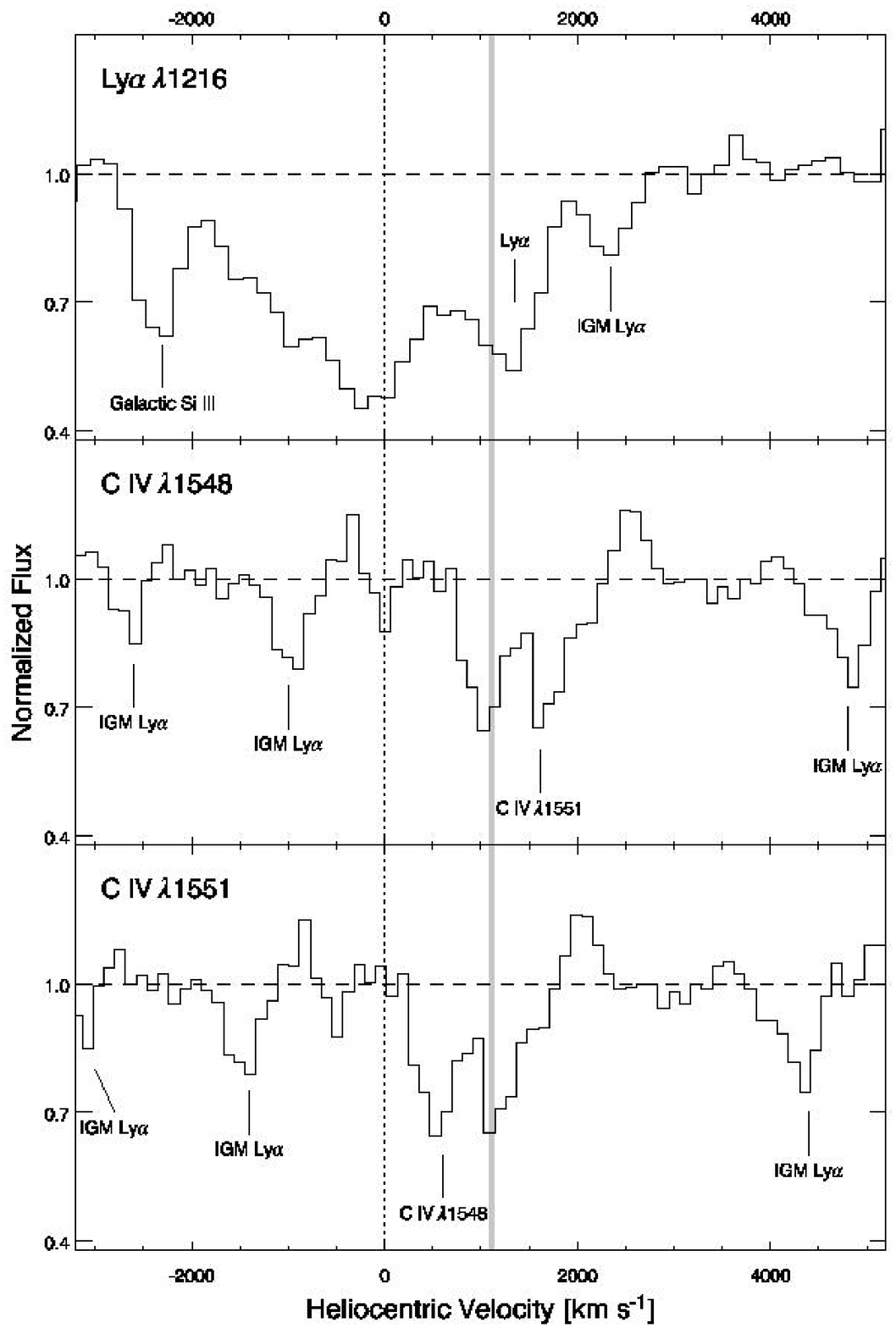}
\end{center}
\caption{\Lya\ ({\em top panel}) and \CIV\ ({\em middle and bottom panels}) absorption features from the HST/STIS snapshot spectrum of \sbs. The best-fit absorber velocity of $\czabs = 1110\pm30$~\kms\ (determined from the \CIV\ lines only) is indicated by the vertical gray bar. Features centered near the dotted vertical line at $v=0$ are Galactic and all other absorption features have been identified and labelled.
\label{fig:qsospec}}
\end{figure}

\begin{figure}[!ht]
\begin{center}
\epsscale{1.00}
\plotone{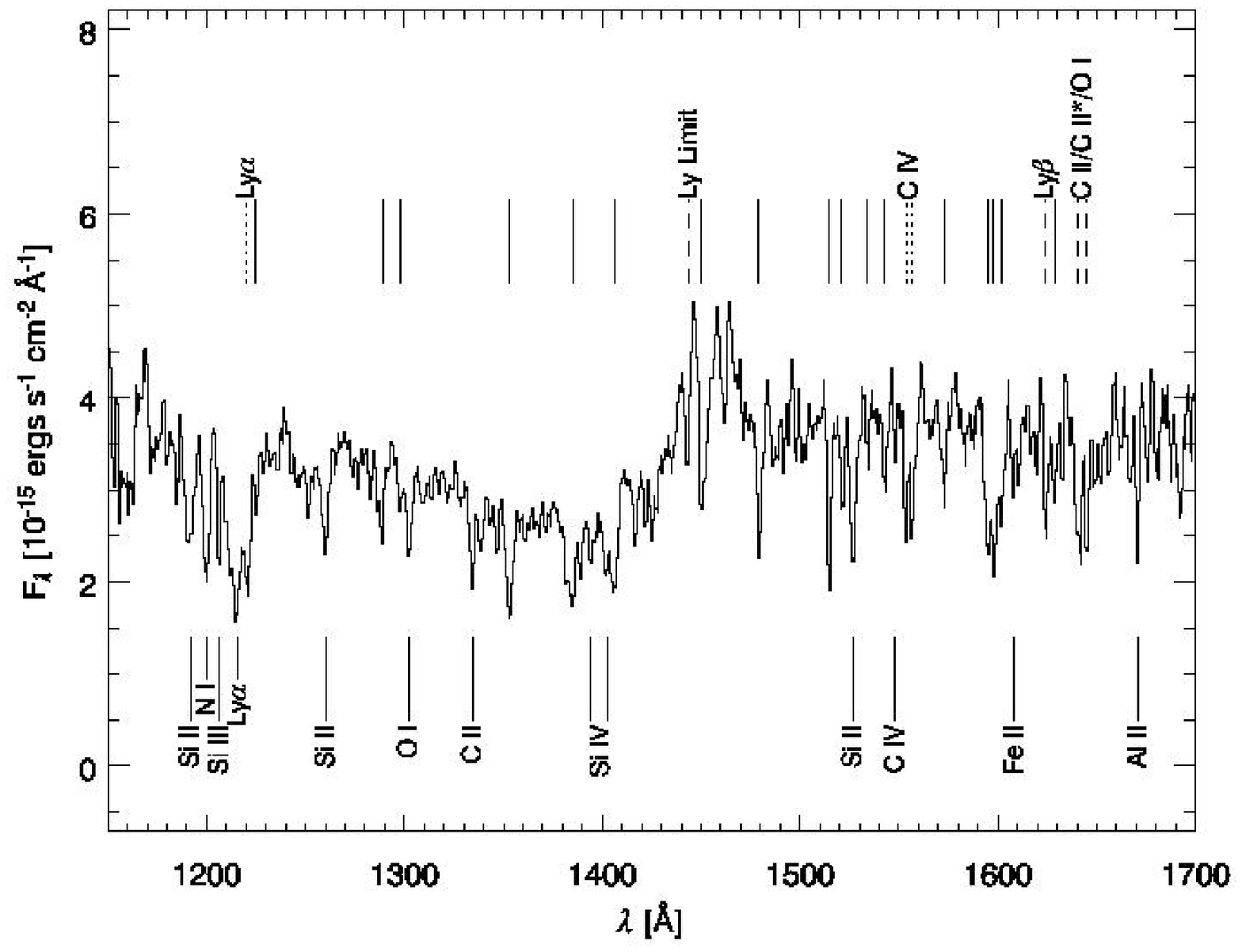}
\end{center}
\caption{The full STIS G140L snapshot spectrum of \sbs. The positions of Galactic lines are indicated with tickmarks below the spectrum and the positions of intergalactic absorption lines are indicated with tickmarks above the spectrum. The \Lya\ and \CIV\ lines at $\czabs = 1110\pm30$~\kms\ (see Fig.~\ref{fig:qsospec}) are shown with dotted tickmarks, lines associated with the LLS at $z=0.583$ are shown with dashed tickmarks, and intergalactic \Lya\ lines at other redshifts are shown with solid tickmarks. The signal-to-noise ratio of the spectrum ranges from $\sim 10$ per resolution element near Galactic \Lya\ to $\sim 5$ per resolution element near Galactic \AlII.
\label{fig:fullqsospec}}
\end{figure}

The positions of Galactic lines in Figure~\ref{fig:fullqsospec} are indicated with tickmarks below the spectrum and the positions of intergalactic absorption lines are indicated with tickmarks above the spectrum. The \Lya\ and \CIV\ lines at $\czabs = 1110\pm30$~\kms\ (see Fig.~\ref{fig:qsospec}) are shown with dotted tickmarks, lines associated with the LLS at $z=0.583$ are shown with dashed tickmarks, and intergalactic \Lya\ lines at other redshifts are shown with solid tickmarks. We have not searched for associated metal lines at the redshifts of the \Lya\ absorbers indicated with solid tickmarks. The broad feature at $\sim 1460$~\AA\ is likely intrinsic \ion{O}{4} 787~\AA\ emission at $z=0.852$. The signal-to-noise ratio of the spectrum ranges from $\sim 10$ per resolution element near Galactic \Lya\ to $\sim 5$ per resolution element near Galactic \AlII, which corresponds to $3\sigma$ equivalent width limits of 360~m\AA\ and 510~m\AA, respectively.

Neither the SDSS spectrum of \sbs\ nor a DIS spectrum taken on 2003 May 2 show \MgII\ absorption at $z=0.58$ with an equivalent width $\gtrsim 50$~m\AA. Luminous ($L > 0.1$--0.3~\Lstar) galaxies are typically found within a projected distance of 50\h~kpc of LLSs with strong \MgII\ absorption \citep{steidel95,steidel98}. Our R-band image of \ic\ (see \S\,\ref{obs:galimg}) does not show a $L \gtrsim 0.3$~\Lstar\ \citep[$m_R \lesssim 21.0$;][]{brown01} galaxy within 50\h~kpc ($6\arcsec$ at $z=0.58$) of \sbs\ that could be responsible for this LLS \citep[the SDSS plates show no galaxy within 50\h~kpc of \sbs\ with $L \gtrsim 0.7$~\Lstar\ in $i'$-band or $L \gtrsim 3$~\Lstar\ in $z'$-band;][]{blanton01}. Our B-band and near-infrared images of \ic\ cannot be used to search for the galaxy responsible for this LLS because they do not have a large enough field of view to cover the \sbs\ sight line. Since the LLS seen in the STIS spectrum of \sbs\ does not produce a strong \MgII\ absorber, it is not surprising that our images do not show a luminous galaxy within a projected distance of 50\h~kpc from the sight line.


\clearpage

\clearpage
\begin{deluxetable}{lc}

\tablecolumns{2}
\tablewidth{0pt}

\tablecaption{\ic\ Optical Emission Line Equivalent Widths
\label{tab:eqw}}

\tablehead{\colhead{Line} & \colhead{\eqw} \\ \colhead{Identification} & \colhead{(\AA)}}

\startdata
$[$\OII$]$~3727\,\AA\dotfill           &   $68\pm5~$   \\
$[$\ion{Ne}{3}$]$~3869\,\AA$\,\dots$   &    $4\pm1$    \\
H$\delta$\dotfill                      &  $2.2\pm0.4$  \\
H$\gamma$\dotfill                      &  $7.4\pm0.5$  \\
$[$\OIII$]$~4363\,\AA\dotfill          &  $1.2\pm0.2$  \\
H$\beta$\dotfill                       &   $28\pm1~$   \\
$[$\OIII$]$~4959\,\AA\dotfill          &   $23\pm1~$   \\
$[$\OIII$]$~5007\,\AA\dotfill          &   $66\pm2~$   \\
\ion{He}{1}~5876\,\AA\dotfill          &    $3\pm1$    \\
$[$\OI$]$~6300\,\AA\dotfill            &  $2.3\pm0.6$  \\
$[$\NII$]$~6548\,\AA\dotfill           &  $7.0\pm0.8$  \\
\Ha\dotfill                            &  $117\pm4~~$  \\
$[$\NII$]$~6584\,\AA\dotfill           &   $20\pm1~$   \\
\ion{He}{1}~6678\,\AA\dotfill          &  $1.6\pm0.3$  \\
$[$\SII$]$~6717\,\AA\dotfill           & $13.0\pm0.6~$ \\
$[$\SII$]$~6731\,\AA\dotfill           & $10.4\pm0.6~$ \\
$[$\ion{Ar}{3}$]$~7136\,\AA\dotfill    &  $4.1\pm0.8$  \\
\enddata

\end{deluxetable}

\end{document}